\documentstyle[amsfonts,12pt]{article}

\def\noi{\noindent}

\def\beq#1{\begin{equation}\label{#1}}
\def\eeq{\end{equation}}

\newcommand{\bear}[1]{\begin{eqnarray}\label{#1}}
\newcommand{\ear}{\end{eqnarray}}

\renewcommand{\theequation}{\arabic{section}.\arabic{equation}}
\catcode`\@=11 \@addtoreset{equation}{section}\catcode`\@=12

\newcommand{\R}{{\mathbb R}}

\newcommand{\btd}{\bigtriangledown}

\newcommand{\apsection}[2]
 {\subsection*{#1. \ #2}
 \renewcommand{\theequation}{#1.\arabic{equation}}
 \setcounter{equation}{0}
 }

\newcommand{\fnm}{\footnotemark}
\newcommand{\fnt}{\footnotetext}

\begin{document}

\begin{center}

\large\bf THICK BRANE WORLD MODEL \\ FROM PERFECT FLUID
\\[15pt]
\normalsize\bf V.D. Ivashchuk\fnm[1]\fnt[1]{ivas@rgs.phys.msu.su},
and V.N. Melnikov\fnm[2]\fnt[2]{mel@rgs.phys.msu.su}
\\[10pt]
\it Center for Gravitation and Fundamental Metrology,
VNIIMS, 3/1 M. Ulyanovoy Str.,
Moscow 117313, Russia  and \\
Institute of Gravitation and Cosmology, People's Friendship
University of Russia, Mikhlukho-Maklaya St. 6,  Moscow 117198,
Russia

\end{center}

\vspace{15pt}

\small\noi

\begin{abstract}

A $(1 +d)$-dimensional thick "brane world" model
with varying $\Lambda$-term is considered.
The model is generalized to the
case of a chain of Ricci-flat internal spaces
when the matter source is an
anisotropic perfect fluid.
The "horizontal" part of potential is obtained
in the Newtonian approximation.
In the multitemporal case (with a $\Lambda$-term)
a set of equations for potentials  is
presented.

\vspace{1cm}

%PACS: 04.20.J, 04.60+n

\vspace{1mm}

%Keywords: thick brane world, perfect fluid

\end{abstract}

\vspace{10cm}

\pagebreak

\normalsize

%%%%%%%%%%%%%%%%%%%%%%%%%%%%%%%%%%%%%%%%%%%%%%%%%%%%%%%%%%%%%%%%
\section{Introduction}
%%%%%%%%%%%%%%%%%%%%%%%%%%%%%%%%%%%%%%%%%%%%%%%%%%%%%%%%%%%%%%%%

An interest in the so-called  "brane world"  models (see
\cite{RubS}-\cite{YuraSht} and refs. therein) has recently
greatly increased due to Refs. \cite{ArDDK,RS2,RS1}. The idea of
a brane world  is rather simple. It is supposed that we are
living on a $(1+3)$-dimensional thin (or thick) layer (3-brane) in
multidimensional space-time and there exists a potential
preventing us from leaving this layer, i.e. gauge and
matter fields are localized on branes whereas gravity "lives" in
the multidimensional bulk. Randall and Sundrum \cite{RS1} suggested
a construction for a confining potential (for an attractive
potential see also \cite{MVV} ), using two symmetric copies of
part of 5-dimensional anti-deSitter space, so that "our"
4-dimensional space-time is a "wedge of the edge". The confining
potential has a $|y|$-type shape, where $y$ is the coordinate of
the extra non-compact fifth dimension.

It should be noted that the main stream of the brane world
studies is related to  thin brane model, although it looks
more natural for this "brane" to be thick rather than thin.

%\fnm[3]\fnt[3]{We note that several rather important questions
%usually are not discussed in publications:
%i) is the "multidimensional world"
%a pure mathematical  construction (remind gauge group fibers
%in a principal bundle), or
%ii) it describes some reality (invisible for us), etc.
%The treatment of the second possibility may lead to a
%situation when scientists will study the subjects that are out of the
%scope of the science itself and, generally speaking, human brain
%facilities. In this case one may confront with a lot of speculations
%belonging rather to science fiction, than to the science itself
%(V.D.I.'s comment).}

A simple $5$-dimensional thick brane model was suggested
in \cite{CsEHS}. In this paper we start with a generalization
of this model to $(d+1)$-dimensional case, i.e. a multidimensional
thick brane model with a $y$-dependent cosmological term is
presented, where $y$ is an extra coordinate (Sec. 2).
(For a review of models with the cosmological term see \cite{SS}.)
Then, we
show that there exist a lot of alternative thick brane models
that may be added to a "brane world collection" after certain
investigations. As an example we consider a brane world model
with a chain of Ricci-flat internal spaces (e.g., compact ones)
and an anisotropic perfect fluid (see \cite{IM5,IM10} and references
therein) as a matter source (Sec. 3). Instead of an $AdS_5$
solution and its thick brane extension, we start with the
Euclidean version of the inflationary solution with perfect
fluid from \cite{IM10} defined on a product of Ricci-flat spaces.
The metric of the  solution  is defined on the  manifold $M = \R
\times M_{1} \times \ldots \times M_{n}$ and has the form
$$g= dy \otimes dy + \sum_{i=1}^{n} \exp[2{\phi^{i}}(y)] g^i,$$
where $(M_i,g^i)$ are Ricci-flat manifolds,
($i = 1, \ldots ,n $; $n \geq 1$) and ($M_{1},g^1)$ is
"our space-time".
We obtain linearized equations for
the potential in the Newtonian approximation
and  present their
("Newton-Yukawa-type") solutions (Sec. 3). In the "$\Lambda$-term"
case, a multitemporal analogue of the set of equations on
the  potentials is
presented (Sec. 2). (For brane-world  models with extra timelike
dimensions see, e.g., \cite{DGS,IK} and references therein.)

%%%%%%%%%%%%%%%%%%%%%%%%%%%%%%%%%%%%%%%%%%%%%%%%%%%%%%%%%%%%%%%%
\section{Thick brane with varying cosmological term}
%%%%%%%%%%%%%%%%%%%%%%%%%%%%%%%%%%%%%%%%%%%%%%%%%%%%%%%%%%%%%%%%

%%%%%%%%%%%%%%%%%%%%%%%%%%%%%%%%%%%%%%%%%%%%%%%%%%%%%%%%%%%%%%%%
\subsection{The model}
%%%%%%%%%%%%%%%%%%%%%%%%%%%%%%%%%%%%%%%%%%%%%%%%%%%%%%%%%%%%%%%%

We first consider a  model governed by the action
\beq{1.1} S=
\int d^Dz \sqrt{|g|} \{R[g]- 2 \Lambda \} + \int d^Dz \sqrt{|g_B|}
\sigma, \eeq
where $g=g_{MN}(z)dz^M\otimes dz^N$ is a metric,
defined on the ($D = 1+d$)-dimensional manifold
\beq{1.2} M^{D} = \R
\times M^d,
\eeq
$z = (z^M) =(x^{\mu}, y)$,  $x^{\mu}$  are
coordinates on $M^d$, $y$ is a coordinate on $\R$, and
\beq{1.2a}
g_B(y) = g_{\mu \nu}(x,y) dx^{\mu} \otimes dx^{\nu}
\eeq
is a $y$-parametrized family of brane metrics defined on
sections $\{y\} \times M^d$ isomorphic to $M^d$, $\mu, \nu = 0,
\ldots, d-1$. We put $|g_B| \equiv |{\rm det} (g_{\mu
\nu}(x,y))| \neq 0.$

The cosmological term $\Lambda$ and the brane tension $\sigma$ are
in general (smooth) functions on $M^D$. We suppose that
these functions only depend on the "extra" dimension,
i.e.
\beq{1.3}
\Lambda = \Lambda(y), \qquad \sigma = \sigma(y).
\eeq

We start with the simplest solutions to the equations of motion
corresponding to the action (\ref{1.1}) defined on the manifold
(\ref{1.2}) with $\Lambda$ and $\sigma$ of the form (\ref{1.3}).
The solution reads
\bear{1.4}
g = e^{2 \phi(y)} g^1 + dy \otimes dy,
\ear
where $g^1$ is a Ricci-flat metric on $M^d$ ($R_{\mu \nu}[g^1] =0$)
and the extra-dimensional ("vertical") potential  $\phi = \phi(y)$
satisfies the following relations:
\bear{1.5}
 d (d-1) (\phi')^2  = - 2 \Lambda, \qquad  2 (d-1) \phi''= \sigma.
\ear
Here and below $\phi'= d \phi/dy = \phi_{,y}$.

Indeed, the Hilbert-Einstein equations  read
 \beq{1.6} R_{MN} -
\frac{1}{2} g_{MN} R  =   T_{MN}^B - \Lambda g_{MN},
\eeq
where the
brane energy-momentum tensor for the block-diagonal metric
$g_{\mu y} =0$ with $g_{yy} =0$ reads (see Appendix A)
\beq{1.7}
(T_{MN}^B)= \frac{\sigma}{2} \left(\begin{array}{cc}
g_{\mu \nu}&0\\
0&0
\end{array}\right).
\eeq
Using the relations for the Ricci-tensor components of $g$ and
the scalar curvature from Appendix B we get the
"fine-tuning" relations (\ref{1.5}).

Let us now consider  the first example: flat pseudo-Euclidean
metric $g^1 = \eta =\eta_{\mu \nu} dx^{\mu} \otimes dx^{\nu}$,
where $(\eta_{\mu \nu}) = {\rm diag}(-1,+1, \ldots, +1)$. Then for
$\sigma = 0$, $\phi(y) = k y$ with constant $k$ satisfying
\beq{1.8}
\Lambda = -d (d-1) k^2/2
\eeq
we get the $D$-dimensional
anti-de Sitter metric in (\ref{1.4}).

For another choice
\beq{1.9} \phi(y) = k |y|, \qquad \sigma = 4
(d-1)k \delta(y)
\eeq
 and $\Lambda$ from (\ref{1.8}) we get a
thin $d$-dimensional brane, attractive for $k > 0$ (for $d=4$ see
\cite{MVV}) and repulsive for $k < 0$ (for $d=4$ see \cite{RS1}).

%%%%%%%%%%%%%%%%%%%%%%%%%%%%%%%%%%%%%%%%%%%%%%%%%%%%%%%%%%%%%%%%
\subsection{Newtonian approximation}
%%%%%%%%%%%%%%%%%%%%%%%%%%%%%%%%%%%%%%%%%%%%%%%%%%%%%%%%%%%%%%%%

Consider small perturbations of  the flat $d$-metric
in the "Newtonian" approximation, i.e., we put
\beq{1.10}
g^1 = - \exp(2v(\vec{x},y)) dt \otimes dt +
(d\vec{x} \otimes d\vec{x})_{d-1},
\eeq
where  $v(\vec{x},y)$ is a small enough "horizontal" part
of the gravitational potential, $t = x^0$ is a time variable.
Here and in what follows
$(d\vec{x} \otimes d\vec{x})_{k} =
\sum_{i=1}^{k} d x^i \otimes d x^i$.

We rewrite Eqs. (\ref{1.6}) in the equivalent form
\beq{1.11}
R_{MN} =  T_{MN}^B + \frac{g_{MN}}{D-2}
(2\Lambda - T),
\eeq
$T = T_{MN}^{B}g^{MN}$. Using the  relation for
the $tt$-component of the Ricci tensor
of the metric (\ref{1.4}) with $g^1$ from (\ref{1.10})
(see Appendix B)
\beq{1.12}
R_{tt}[g] = e^{2(\phi +v)}[e^{-2\phi} \Delta v +  \phi'' +
d(\phi')^2 + v'' + (d+1)v'\phi'  + O(v^2)],
\eeq
we get from  (\ref{1.11}) and (\ref{1.5})
a linearized equation for small $v$
\beq{1.13}
\Delta v +  e^{2\phi}[v'' + (d+1) \phi' v'] = 0,
\eeq
where $\Delta $ is the Laplace operator on $\R^{d-1}$.

Eq. (\ref{1.13}) can be easily solved by
separation of variables, i.e. by seeking solutions as
superposition of monoms:   $v = v_1(\vec{x}) v_2(y)$. We are
interested in spherically symmetric solutions with a certain
behaviour at infinity $|\vec{x}| \to \infty$. The solution reads
\beq{1.14}
v =  \frac{GM}{r^{\bar d}} + \int_{0}^{+\infty } d m
\rho(m,y) \frac{1}{r^{\bar d}} \exp(-m r), \eeq
 where ${\bar d} =d -2$ and
\beq{1.15}
\rho'' + (d+1) \phi' \rho' + m^2 e^{-2\phi}
\rho = 0.
\eeq
$G$ is the ($d$-dimensional) gravitational constant,
$M$ is the mass and $m$ is a spectral parameter. We restrict
ourselves to the "Yukawa-type" part of the general solution.
We thus consider a  superposition of the Newtonian
potential and the generalized Yukawa-type one.

%%%%%%%%%%%%%%%%%%%%%%%%%%%%%%%%%%%%%%%%%%%%%%%%%%%%%%%%%
\subsection{Multitemporal generalization}
%%%%%%%%%%%%%%%%%%%%%%%%%%%%%%%%%%%%%%%%%%%%%%%%%%%%%%%%%%

Consider a multitemporal generalization of the metric
(\ref{1.10}), i.e.
\beq{2.1} g^1 = - \sum_{i =1}^{n}
\exp(2v_i(\vec{x},y)) dt_i \otimes dt_i + (d\vec{x} \otimes
d\vec{x})_{d-n},
\eeq
where  $v_i(\vec{x},y)$ is a small enough
part of the gravitational potential corresponding to $t_i$, $i = 1,
\ldots, n$, $n >1$. Using the relations for the Ricci-tensor
components (see Appendix B)
\beq{2.2} R_{t_it_i}[g] =  e^{2(\phi
+v_i)} [ e^{-2\phi}\Delta v_i  + \phi'' +  d (\phi')^2 + v''_i +
\phi' (d v_i' + \sum_{j=1}^{n} v_j') + O(v^2)],
\eeq
we get from
Eqs. (\ref{1.11}) and (\ref{1.5}) the following relations:
\beq{2.3}
\Delta v_i +  e^{2\phi}[v''_i +  \phi' (d v_i' +
\sum_{j=1}^{n} v_j')] = 0,
\eeq
$i =1, \ldots, n$. Here $\Delta$
is the Laplace operator on  $\R^{d-n}$. For $n= 1$ we get  Eq.
(\ref{1.13}). Eqs. (\ref{2.3}) can be easily solved in the symmetric
case, when $v_i = v$. In this case we get the relations (\ref{1.14})
with ${\bar d} = d -1 -n$ and  (\ref{1.15})  with $d$
replaced by $d+n$. We note that in the vacuum case $\phi = 0$
the multitemporal analogs of the Schwarzschild
and Tangherlini solutions
>from \cite{IMT} give us exact solutions to the field equations.

%%%%%%%%%%%%%%%%%%%%%%%%%%%%%%%%%%%%%%%%%%%%%%%%%%%%%%%%%%%%%%%%%%
\section{Thick brane with perfect fluid on a product of $n+1$ spaces}
%%%%%%%%%%%%%%%%%%%%%%%%%%%%%%%%%%%%%%%%%%%%%%%%%%%%%%%%%%%%%%%%%%

Consider a generalization of the thick brane solution from
the previous section to the case of $n -1$ Ricci-flat spaces and
perfect fluid as a matter source (see \cite{IM5,IM10} and references
therein).

We take the metric
\begin{equation}
\label{3.1}
g= dy \otimes dy + \sum_{i=1}^{n} \exp[2{\phi^{i}}(y)] g^i,
\end{equation}
defined on the  manifold
\begin{equation} \label{3.2}
M = \R \times M_{1} \times \ldots \times M_{n},
\end{equation}
where the manifolds $M_{i}$ with the metrics $g^i$ are
Ricci-flat spaces of dimensions $d_{i}$, $i = 1, \ldots ,n $;
$n \geq 1$.
One of the spaces, say ($M_{1},g^1)$, is by convention "our space-time"
and other spaces are "internal".
Consider the Einstein equations
\beq{3.3}
R^{M}_{N}-\frac{1}{2}\delta^{M}_{N}R = \kappa^{2}T^{M}_{N},
\eeq
where $\kappa^{2}$ is the gravitational constant and
the energy-momentum tensor is adopted in the form
\begin{equation}
\label{3.4}
(T^{M}_{N})= {\rm diag}( p_y,  p_{1} \delta^{m_{1}}_{k_{1}},
\ldots , p_{n} \delta^{m_{n}}_{k_{n}}).
\end{equation}
Here $p_y = p_y(y)$ is the pressure in the 1-dimensional $y$-space $\R$
and $p_i = p_i(y)$ is the pressure in $M_i$,
$i = 1, \ldots, n $.

Using relations for Einstein tensor components
$E^M_N =R^M_N - \frac{1}{2}\delta^M_N R$
for the metric (\ref{3.1}) (see Appendix C),
we obtain from Einstein Eqs. (\ref{3.3})
\bear{3.5}
p_y = - \frac{1}{2\kappa^{2}} G_{ij} \phi^i_{,y} \phi^j_{,y}, \\
\label{3.6}
p_i = - p_y
- \frac{1}{\kappa^{2} d_i}G_{ij} (\phi^j_{,yy} +
\phi^j_{,y} d_{k} \phi^k_{,y}),
\ear
$i = 1, \ldots, n $. Here
\begin{equation} \label{3.7}
G_{ij} = d_{i} \delta_{ij} - d_{i}d_{j},
\end{equation}
are components of the minisuperspace metric,
$i,j = 1, \ldots, n$ \cite{IMZ}. The minisupermetric
has a pseudo-Euclidean signature.

The pressures may be decomposed into "bulk"
and "brane" parts
\beq{3.7c}
p_i = p_i^{bulk}  + p_i^{br},
\eeq
where by definition
\bear{3.7a}
p_i^{bulk} = - p_y
- \frac{1}{\kappa^{2} d_i}G_{ij} \phi^j_{,y} d_{k}\phi^k_{,y},
\\  \label{3.7b}
p_i^{br} = - \frac{1}{\kappa^{2}d_i}G_{ij} \phi^j_{,yy},
\ear $
i = 1, \ldots, n$.
Defining  $p_y^{bulk} = p_y$, $p_y^{br} = 0$,
we decompose the stress-energy tensor into a sum of two
components: $T^{M}_{N} =T^{bulk,M}_{N} + T^{br,M}_{N}$, where \\
$(T^{bulk,M}_{N})= {\rm diag}( p_y,  p^{bulk}_{1}
\delta^{m_{1}}_{k_{1}}, \ldots , p^{bulk}_{n}
\delta^{m_{n}}_{k_{n}})$, $(T^{br,M}_{N})= {\rm diag}( 0,
p^{br}_{1} \delta^{m_{1}}_{k_{1}}, \ldots , p^{br}_{n}
\delta^{m_{n}}_{k_{n}})$. As we shall see below these
decompositions will be justified by the thin brane limit.

{\bf One-function dependence.}
Consider  the following ansatz
\beq{3.8}
\phi^i(y) = - \kappa \frac{u^i}{\sqrt{-<u,u>}} f(y),
\eeq
where $f(y)$ is a smooth function, $u^i = G^{ij}u_j$, $u = (u_i) \in
\R^{n}$,
\beq{3.8a}
<u,v> = G^{ij} u_i v_j
\eeq
is a scalar product on $\R^n$, $u,v \in \R^{n}$,
and
\begin{equation}  \label{3.9a}
G^{ij} = \frac{\delta^{ij}}{d_{i}} + \frac{1}{2-D}
\end{equation}
are components of the matrix inverse to  $(G_{ij})$
($D = 1 + \sum_{i=1}^{n} d_i$) \cite{IMZ}.
Here we suppose that
\beq{3.9}
<u,u> < 0.
\eeq
As we shall see below, this condition is satisfied for isotropic case
when all pressures are equal: $p_i =p$, $i = 1, \ldots, n$.

For the pressures (\ref{3.7a}) and (\ref{3.7b})  we obtain
in the special case (\ref{3.8})
\bear{3.10a}
p_i^{bulk} =
\left(\frac{<u^{\Lambda},u> u_i}{<u,u> d_i} - 1\right) p_y ,
\qquad (f'(y))^2 = 2 p_y(y)
\\  \label{3.10b}
p_i^{br} = \frac{u_i}{\kappa d_i \sqrt{-<u,u>}} f''.
\ear
Here the vector
\beq{3.r3}
u^{\Lambda}_i = 2 d_i,
\eeq
$i = 1, \dots, n$, corresponds to the cosmological constant case.

Since  the relations  for the metric and pressures are invariant
under the replacement $u \mapsto \lambda u$ ($\lambda > 0$)
we may normalize $u = (u_i)$ by the condition
\beq{3.11}
<u^{\Lambda}, u> = <u,u>.
\eeq

In this case Eq. (\ref{3.10a}) reads
\beq{3.12}
p_i^{bulk} = \left(\frac{u_i}{d_i} - 1 \right) p_y,
\eeq
$i = 1, \dots, n$.

We note that for $f(y) = \sqrt{2p_y} y$ the solution
is nothing else but a Euclidean version of the (exponential)
inflationary solutions from \cite{IM10} (for $u = u^{\Lambda}$ see
also \cite{BIMZ}).
For $n=1$ one get $u = u^{\Lambda}$ due to
the normalization condition (\ref{3.11}), thus the perfect
fluid generalizations are non-trivial  only for $n > 1$.

{\bf Isotropic case.}
For $u_i = h d_i$, we get from (\ref{3.11})  $h=2$,
i.e. we are led to cosmological constant case:
$u^{\Lambda} = u$ with
\beq{3.r2}
u^i = \frac{2}{2 - D}, \qquad
<u,u> =  - 4 \frac{D - 1}{D - 2} < 0,
\eeq
$i = 1, \dots, n$. Thus the
restriction (\ref{3.9}) is satisfied identically.
In this case we get
\beq{3.8b}
\phi^i(y) =  \frac{\kappa}{\sqrt{(D-1)(D-2)}} f(y),
\eeq
and
\beq{3.13}
p_i^{br} = \frac{\sqrt{D-2}}{\kappa \sqrt{D-1}} f'',
\qquad
p_i^{bulk} = p_y = (f'(y))^2 /2,
\eeq
$i = 1, \dots, n$. Since all scale factors are equal,
one can reduce the isotropic case to the 1-space case
making the redefinition $g^1 + \dots g^n = \bar{g}^1$.
The varying cosmological constant reads:
$\Lambda(y) = - \kappa^2 p_y(y)$.
For  $n =1$  we get the relations (\ref{1.5}) with $2p_1^{br} = \sigma$.

{\bf Thin brane.} Consider  a special solution  with
\beq{3.13a}
\phi^i(y) = \sqrt{2p_y} |y|,
\eeq
$p_y > 0$. In this case we get
a "thin brane" with the pressures
\beq{3.14}
p_i^{br} = \frac{u_i \sqrt{2p_y}}{\kappa d_i \sqrt{-<u,u>}} 2 \delta(y),
\eeq
$i = 1, \dots, n$.  We see that the thin-brane tension is
proportional to a square route of the pressure in the $y$-dimension.

{\bf Newtonian approximation.}
Using a relation for $R_{tt}[g]$ from Appendix B in
the multispace case with $g^1$ from (\ref{1.10}),
we get a modification of the Eq. (\ref{1.13})
to the perfect-fluid case with $(n-1)$ internal
Ricci-flat spaces
\beq{3.16}
\Delta v +  e^{2\phi^1} \{ v'' +
[(\phi^1)' + \sum_{i=1}^{n} d_i (\phi^i)'] v' \} = 0,
\eeq
where $\Delta $ is the Laplace operator on $\R^{d_1 - 1}$.
Here all information about the perfect fluid
and the internal spaces is hidden in the behavior of functions
$\phi^i(u)$. The formal solution (\ref{3.16})
coincides with that of (\ref{1.14}) ($d = d_1$)
but the equation for the density (\ref{1.15})
should be modified as follows
\beq{3.17}
\rho'' + [(\phi^1)' + \sum_{i=1}^{n} d_i (\phi^i)']
\rho' + m^2 e^{-2\phi^1} \rho = 0.
\eeq

\section{Conclusions and discussion}

Here  we have considered a generalization of the $5$-dimensional thick
brane model with a $y$-dependent cosmological term from
\cite{CsEHS} to the multidimensional case  (Section 2) and to
a multifactor product-space case, when the anisotropic perfect fluid
(Sec. 3) is adopted as a matter source. These  models have
thin brane limits. In both cases we have obtained linearized equations
for small "horizontal" potentials and their Newton-Yukawa-type
solutions. In the "$\Lambda$-term" case we have also considered
a multitemporal generalization and a set of linearized equations
for potentials is written (it is solved in the symmetric case). In
the perfect fluid case we have suggested two brane world models:  (i)
a general one, with more or less arbitrary $y$-dependent scale
factors and pressures (decomposed into "perfect" and "brane"
parts):  (ii) a special model  governed by one function
$f(y)$ and the anisotropy parameters $u_i$. In the thin brane limit
the latter coincides for $y >0$ with the Wick-rotated
inflationary solution from \cite{IM10} (and symmetrized
with respect to reflection in $y$).

A further consideration needs a globally consistent treatment of
linearized gravity and exact solutions. Some special
smeared (or regularized) thin brane solutions may be also
considered (e.g., with a cosmological type metric $g^1$ ).

%\fnm[4]\fnt[4]{We should be also prepared to a situation in future when
%the "brane world" approach will either fail,
%being in contradiction with experiment and observations,
%or may lead to a situation when a huge number of models
%with large domains of free parameters will be allowed
%(e.g. as it takes place in Brance-Dicke model). Of course,
%such "destiny" of brane world paradigm will allow to
%theoreticians to write a lot of papers but will
%not explain to experimentalists what is the right theory.
%Here we may need a lot in propositions {\em a l\'a } "no-go"
%theorems \cite{GKL} or some sort of censorship analysis
%in order to close possible "ways to nowhere"
%(V.D.I.).}

\begin{center}
{\bf Acknowledgments}
\end{center}

This work was supported in part by
the Russian Ministry for Science and Technology, Russian Foundation for
Basic Research, and project SEE.
One of the authors (V.D.I.) thanks A.A. Starobinsky and V.A. Rubakov
for discussions, stimulating the interest to the thick brane model.

\section*{Appendix}

\apsection{A} {Energy-momentum tensor for thick brane}

Consider the thick brane part of the action
\beq{A1.1}
S_B[g]=  \int d^Dz \sqrt{|g_B|} \sigma(z)
\eeq
where $|g_B| = |{\rm det} g_{\mu \nu}| \neq 0$
($\mu, \nu = 0, \ldots, d-1$)
and the metric $g=g_{MN}(z)dz^M\otimes dz^N$ is defined
on the $(D = 1+d)$-dimensional manifold (\ref{1.2}).

We take the following representation of the metric
\beq{A1.2}
(g_{MN})=
\left(\begin{array}{cc}
g_{\mu \nu}^B& n_{\mu}\\
n_{\nu}    &n_{\rho} n^{\rho} + b
\end{array}\right),
\eeq
where $b = {\rm det}( g_{M N})/{\rm det} (g_{\mu \nu}) \neq 0$.
For the inverse matrix we get
\beq{A1.3}
(g^{MN})=
\left(\begin{array}{cc}
g^{\mu \nu}_B + b^{-1} n^{\mu} n^{\nu}& -n_{\mu}b^{-1}\\
-n_{\nu}b^{-1}  & b^{-1}
\end{array}\right),
\eeq
where $(g^{\mu \nu}_B) =  (g_{\mu \nu})^{-1}$
and $n^{\mu} = g^{\mu \nu}_B n_{\nu}$.

Variation of the action (\ref{A1.1})
\beq{A1.4}
\delta S_B[g]=  (- 1/2) \int d^Dz \sqrt{|g_B|} \sigma
[g_{\mu \nu} \delta g^{\mu \nu} + 2 n_{\mu} \delta g^{\mu y}
+ n_{\rho} n^{\rho} \delta g^{yy}]
\eeq
implies  the energy-momentum  tensor for the thick brane
\beq{A1.5}
(T_{MN}^B)=
\frac{\sigma}{2} \sqrt{\frac{|g|}{|g_B|}}
\left(\begin{array}{cc}
g_{\mu \nu}& n_{\mu}\\
n_{\nu}    &n_{\rho} n^{\rho}
\end{array}\right).
\eeq
We note that ${\rm det} (T_{M N}^B) = 0$.

\apsection{B} {Ricci-tensor components}

{\bf $(1+d)$-dimensional case.}
Nonzero Ricci tensor components
for the metric (\ref{1.4}) are
\bear{A2.1.1}
R_{\mu \nu}[g]  = -  g_{\mu \nu}  [\phi'' + d (\phi')^2],
\\ \label{A2.1.1a}
R_{yy}[g]  =  - d [\phi'' + (\phi')^2].
\ear
The scalar curvature for (\ref{1.4}) is
\beq{A2.1.2}
  R[g] =  - 2d \phi'' - d(d+1)(\phi')^2.
\eeq

For the metric (\ref{1.4})
with $g^1$ from (\ref{1.10}) we get
\beq{A2.1.4}
R_{tt}[g] =  e^{2(\phi +v)} \{
e^{-2\phi}[\Delta v + (\btd v)^2] +
\phi'' + v'' + (\phi' + v')(d \phi' + v') \}.
\eeq

{\bf Multitemporal case.}
For the metric (\ref{1.4})
with $g^1$ from (\ref{2.1}) we obtain
\beq{A2.2.2}
R_{t_it_i}[g] =  e^{2(\phi +v_i)} \{
e^{-2\phi}[\Delta v_i + \sum_{j =1}^{n} \btd v_i \btd v_j ] +
\phi'' + v_i'' + (d -n) \phi' (\phi' + v'_i)
+ (\phi' + v'_i)\sum_{j =1}^{n}(\phi' + v'_j) \},
\eeq
$i =1, \ldots, n$. Here $\Delta$ is the Laplace operator
on  $\R^{d-n}$.

{\bf Product of $n$ spaces.}
Consider the product space
metric (\ref{3.1}) with $g^1$ from (\ref{1.10})
and $d_1 = d$. Calculations give us
\beq{A2.3.1}
R_{tt}[g] =  e^{2(\phi^1 +v)} \{
e^{-2\phi^1}[\Delta v + (\btd v)^2] +
(\phi^1)'' + v'' + [(\phi^1)' + v'] [v' +
\sum_{i=1}^{n} d_i (\phi^i)') \},
\eeq
$\Delta$ being the Laplace operator on  $\R^{d_1-1}$.

\apsection{C} {Einstein tensor}

Let us  present expressions for the Einstein tensor
$E_{MN} =  R_{MN}-\frac{1}{2}g_{MN}R$
corresponding to the metric (\ref{3.1})
(see Appendix in \cite{IMBil})
\bear{A3.1}
E_{uu} = - L \exp( - \gamma_0), \\
\label{A3.2}
E_{m_{i} n_{i}} = -
 \frac{1}{d_i} g_{m_i n_i}^i
 \left(\frac{d}{du} \frac{\partial L}{\partial (\phi^i)'} -
 \frac{\partial L}{\partial \phi^i}\right) \exp(2 \phi^i  - \gamma_0)
 \ear
$i=1,\ldots,n$, where
\beq{A3.3}
L = \frac{1}{2} \exp(\gamma_0) G_{ij} (\phi^i)' (\phi^j)'
\eeq
is Lagrangian
and $\gamma_0 = \sum_{i = 1}^{n} d_i \phi^i$.


\small
\begin{thebibliography}{99}

\bibitem{RubS}
V.A. Rubakov and M.E. Shaposhnikov,
{\it Phys. Lett.} {\bf B 125}, 139 (1983).

\bibitem{Akama}
K. Akama, in {\it Gauge Theory and Gravitation} ed. by
K. Kikkawa, N. Nakanishi, and H. Nariai
(Springer-Verlag, 1983).

\bibitem{Vis}
M. Visser,
{\it Phys. Lett.} {\bf B 159}, 22 (1985).
\bibitem{Sq}
E.J. Squires, {\it Phys. Lett.} {\bf B 167}, 286 (1986).

\bibitem{GibW}
G.W. Gibbons and D.L. Wiltshire,
{\it Nucl. Phys.} {\bf B 287}, 717 (1987).

\bibitem{Gog}
M. Gogberashvili, {\it Phys. Lett.}  {B 484}, 124  (2000) (see
also hep-ph/9812296; hep-ph/9812365; hep-ph/9904383;
hep-ph/9908347).

\bibitem{BDL}
P. Bin\'etruy, C. Deffayet and D. Langlois,
{\it Nucl. Phys.}  {\bf B 565}, 269 (2000); hep-th/9905012.

\bibitem{ArDDK}
N. Arkani-Hamed, S. Dimopoulos, G. Dvali and N. Kaloper,
{\it Phys. Rev. Lett.} {\bf 84}, 586 (2000); hep-th/9907209.

\bibitem{RS2}
L. Randall and R. Sundrum, A large mass hierarchy from a small
extra dimension, {\it Phys. Rev. Lett.}  {\bf 83}, 3370 (1999);
hep-ph/9905221.

\bibitem{RS1}
L. Randall and R. Sundrum,
An alternative to compactification,
{\it Phys. Rev. Lett.}  {\bf 83}, 4690 (1999); hep-th/9906064.

\bibitem{Kog}
I.I. Kogan, S. Mouslopoulos, A. Papazoglou, G.G. Ross, A three
three-brane universe: New phenomenology for the new millennium?
{\it Nucl. Phys.} {\bf B 584}, 313 (2000); hep-ph/9912552.

\bibitem{DGS}
G. Dvali, G. Gabadadze and G. Senjanovi\'c, Constraints on extra
time dimensions, hep-ph/9910207.

\bibitem{Maeda}
T. Shiromizu, K. Maeda and M. Sasaki, {\it Phys. Rev.} {\bf D
62},  024012 (2000).

\bibitem{CsEHS}
C. Cs\'aki,  J. Erlich, T.J. Hollowood and Y. Shirman Universal
Aspects of Gravity Localized on Thick Branes, {\it Nucl.
Phys.} {\bf B 581}, 309 (2000); hep-th/0001033.

\bibitem{NO}
S. Nojiri and S.D. Odintsov, {\it Phys. Lett.} {\bf B 484} 119
(2000); hep-th/0004097; \\
S. Nojiri, S.D. Odintsov and S. Zerbini, {\it Phys. Rev.} {\bf D
62}, 064006  (2000), hep-th/0001192;
for a review, see S. Nojiri and S.D. Odintsov, hep-th/0105160.

\bibitem{HHR}
S.W. Hawking, T. Hertog and H.S. Reall,
{\it Phys. Rev.} {\bf D 62} (2000) 043501; hep-th/0003052.

\bibitem{GRS}
R. Gregory, V.A. Rubakov and S.M. Sibiryakov, Opening up extra
dimensions at ultra-large scales, {\it Phys. Rev. Lett.} {\bf
84}, 5928 (2000), hep-th/0002072.
%Brane-worlds: the gravity of
%escaping matter, hep-th/0003109.

\bibitem{MVV}
W. M\"uck, K.S. Viswanatan and I.V. Volovich,
Geodesic and Newton's Law in Brane Backgrounds,
{\it Phys. Rev.} {\bf D 62} 105019 (2000); hep-th/0002132.

\bibitem{AIMVV}
I.Ya. Aref'eva, M.G. Ivanov, W. M\"uck, K.S. Viswanatan and I.V.
Volovich, Consistent Linearized Gravity in Brane Backgrounds,
{\it Nucl. Phys.} {\bf B 590}, 273 (2000); hep-th/0004114.

\bibitem{GKL}
G. Gibbons, R. Kallosh and  A. Linde, Brane World Sum Rules,
{\it JHEP} {\bf 0101}, 022 (2001); hep-th/0011225.

\bibitem{IK}
A. Iglesias and Z. Kakushadze,
Time-like Extra Dimensions without Tachyons or Ghosts,
hep-th/0012049.

\bibitem{YuraSht}
Yu.V. Shtanov, Flat vacuum branes without fine tuning,
hep-ph/0108211.

\bibitem{SS}
V. Sahni and A. Starobinsky,  The case for a Positive
Cosmological Lambda-term,
{\it Int. J. Mod. Phys.}, {\bf D 9}, 373-444 (2000).

\bibitem{IMT}
V.D. Ivashchuk  and  V.N. Melnikov,
Multi-temporal Generalization
of the Tangherlini Solution, {\it Class. Quantum Grav.} {\bf 11},
1793-1805 (1994);
Multitemporal Generalization of the Schwarzschild
Solution,  {\it Int. J. Mod. Phys.}, {\bf D 4}, No. 2, 167-173 (1995).

\bibitem{IM5}
V.D. Ivashchuk and V.N. Melnikov, Multidimensional cosmology with
$m$-component perfect fluid, gr-qc/9403063; {\it Int. J. Mod.
Phys.} {\bf D 3}, 795-811 (1994).

\bibitem{IM10}
V.D. Ivashchuk and V.N. Melnikov, Multidimensional Classical and
Quantum Cosmology with Perfect Fluid, {\it Grav.
Cosmol.} {\bf 1}, No 2, 133-148 (1995); Inflationary Solutions
in Multidimensional Cosmology with Pefect Fluid, preprint
CBPF-NF-034/95, (Rio de Janeiro) (1995).

\bibitem{BIMZ}
U. Bleyer, V.D. Ivashchuk, V.N. Melnikov and A.I. Zhuk,
Multidimensional Classical and Quantum Wormholes in Models with
Cosmological Constant, {\it Nucl. Phys.}, {\bf B 429}, 177-204,
(1994).

\bibitem{IMZ}
V.D. Ivashchuk, V.N. Melnikov and A.I. Zhuk, On Wheeler-DeWitt
Equation in Multidimensional Cosmology, {\it Nuovo  Cimento},
{\bf B 104}, No 5, 575-582 (1989).

\bibitem{IMBil}
V.D. Ivashchuk and V.N. Melnikov, Billiard Representation for
Multidimensional Cosmology with Multicomponent Perfect Fluid near
the Singularity, {\it Class. Quantum Grav.}, {\bf 12}, No 3,
809-826 (1995).


\end{thebibliography}
\end{document}